\documentclass[aps,pre,twocolumn,groupedaddress,showpacs]{revtex4-1}

\usepackage{color}
\usepackage{graphicx}
\newcommand{\mM}[0]{\mathcal{M}}
\newcommand{\mN}[0]{\mathcal{N}}
\newcommand{\mD}[0]{\mathcal{D}}
\newcommand{\mF}[0]{\mathcal{F}}
\newcommand{\mP}[0]{\mathcal{P}}
\newcommand{\mA}[0]{\mathcal{A}}
\newcommand{\eqn}[1]{
\begin{eqnarray}
	#1
\end{eqnarray}}

\begin{document}


\title{Nonequilibrium Equalities in Absolutely Irreversible Processes}


\author{Y\^uto Murashita}
\author{Ken Funo}
\author{Masahito Ueda}
\affiliation{Department of Physics, University of Tokyo, 7-3-1 Hongo, Bunkyo-ku, Tokyo 113-8654, Japan}


\date{\today}

\begin{abstract}
We generalize nonequilibrium integral equalities to situations involving {\it absolutely irreversible} processes for which the forward-path probability vanishes and the entropy production diverges, rendering conventional integral fluctuation theorems inapplicable.
We identify the mathematical origins of {\it absolute irreversibility} as the singularity of probability measure.
We demonstrate the validity of the obtained equalities for several models.
\end{abstract}


\maketitle


\section{Introduction}
The last two decades have witnessed remarkable progress in 
nonequilibrium statistical mechanics \cite{ECM93, ES94, GC95, Kur98, Ls99, Mae99, LRB00, Jar97Apr, Jar97Nov, Jar00, Cro98, Cro99, Cro00, HS01, Jar04, KPvB07, PvBK09, EvB11, SU10, Sag11, SU12, SU12E, HV10, MT11, Sei12, FWU13, LDSe02, LDSe02, WSMe02, DCPR05, CRJe05, CRJe05, GVNe11, TSUe10}.
Jarzynski established an integral nonequilibrium equality based on the Hamiltonian dynamics \cite{Jar97Apr} and subsequently generalized it to a broader class of situations in nonequilibrium statistical mechanics \cite{Jar97Nov, Jar00}.
Crooks gave a general proof of the Jarzynski equality and fluctuation theorems in stochastic systems, based on a technique of comparing a thermodynamic process and its time-reversed one \cite{Cro98, Cro99, Cro00}.
Recently, the Jarzynski equality has been generalized to situations under feedback control \cite{SU10, Sag11}.
Various types of the Jarzynski equality can be derived in stochastic systems by an appropriate choice of the reference probability  \cite{Sei12}.

Possible applications of nonequilibrium fluctuation equalities range from physics to biology \cite{LDSe02, WSMe02, DCPR05, CRJe05, GVNe11, TSUe10}.
For example, the free-energy landscape of a DNA can be surveyed through nonequilibrium experiments \cite{GVNe11} using the Hummer-Szabo equality \cite{HS01}.
Moreover, feedback control based on information processing as in Ref.~\cite{TSUe10} can be utilized to manipulate nanomachines subject to large thermal fluctuations such as biological nanomachines.

Although nonequilibrium equalities have wide applications, they cannot be applied to such important processes as free expansion, processes starting from local equilibrium, and feedback control with error-free measurements.
In fact, the Jarzynski equality is known to be inapplicable to free expansion of an ideal gas \cite{Sun05, LG05, Gro05Aug, Jar05, Gro05Sep}, because there are some regions in which the forward-path probability vanishes and the backward-path probability does not \cite{HV10}.
A similar difficulty also arises when the initial probability distribution is confined to a restricted region in phase space (e.g. when the process starts from a local equilibrium state).
In these cases, the Crooks fluctuation theorem \cite{Cro99}
\eqn{\label{CFT}
	\frac{\mP^\dag[\Gamma^\dag]}{\mP[\Gamma]} = e^{-\sigma}
}
leads to a negatively divergent entropy production,
where $\Gamma$, $\sigma$, $\mP$, and $^\dag$ represent a path in phase space, entropy production, path probability, and the time reversal, respectively.
In Ref. \cite{HV10}, it has been argued that this divergence of the exponentiated entropy production is circumvented at the level of the detailed fluctuation theorem by exchanging the denominator and numerator:
\eqn{
	\frac{\mP[\Gamma]}{\mP^\dag[\Gamma^\dag]}=e^{\sigma}.
}
In this formula, $e^{\sigma}$ remains finite even when $\sigma$ diverges negatively.
However, the absence of the divergent entropy is implicitly assumed when integral fluctuation equalities are derived in Ref. \cite{HV10}.
Situation-specific modifications are needed to derive rather unusual integral fluctuation equalities in the above-described situations.
Moreover, integral nonequilibrium equalities under error-free measurements have been elusive because the error-free property forbids some forward paths, again leading to divergent entropy production.
In fact, these situations are explicitly excluded when integral fluctuation theorems are derived in Ref. \cite{SU12}.

In this paper, we introduce a concept of {\it absolute irreversibility} as a new class of irreversibility which encompasses the entire range of those irreversible situations to which conventional fluctuation theorems cannot apply, including in the above situations.
In ordinary thermodynamic evolutions, irreversible dynamics are usually stochastically reversible in the sense that both the original and reference (e.g. time-reversed) path probabilities are nonzero.
However, when the original path probability vanishes, the process is not even stochastically reversible.
We call these processes absolutely irreversible because
absolute irreversibility is the direct cause of the divergent entropy.
We identify the mathematical origin of absolute irreversibility as the singularity of the reference probability measure.
This insight, together with Lebesgue's decomposition theorem, enables us to deal with the hitherto excluded situations and derive new nonequilibrium equalities in a unified manner.
What is remarkable and rather unusual is the fact that such a purely mathematical notion corresponds exactly to the physical difficulty of the conventional fluctuation theorems.

This paper is organized as follows.
In Sec. II, we introduce a concept of absolute irreversibility and derive nonequilibrium equalities without feedback control.
In Sec. III, these equalities are verified in free expansion analytically and in overdamped Langevin systems numerically.
In Sec. IV, we generalize the obtained equalities to cases with measurements and feedback control and demonstrate the validity of the obtained equalities in simple models.
In Sec. V, we conclude this paper.

\section{Absolute Irreversibility and Nonequilibrium Equalities}
\subsection{Absolute Irreversibility}
\begin{figure}
\begin{center}
\includegraphics[width = 0.7\columnwidth]{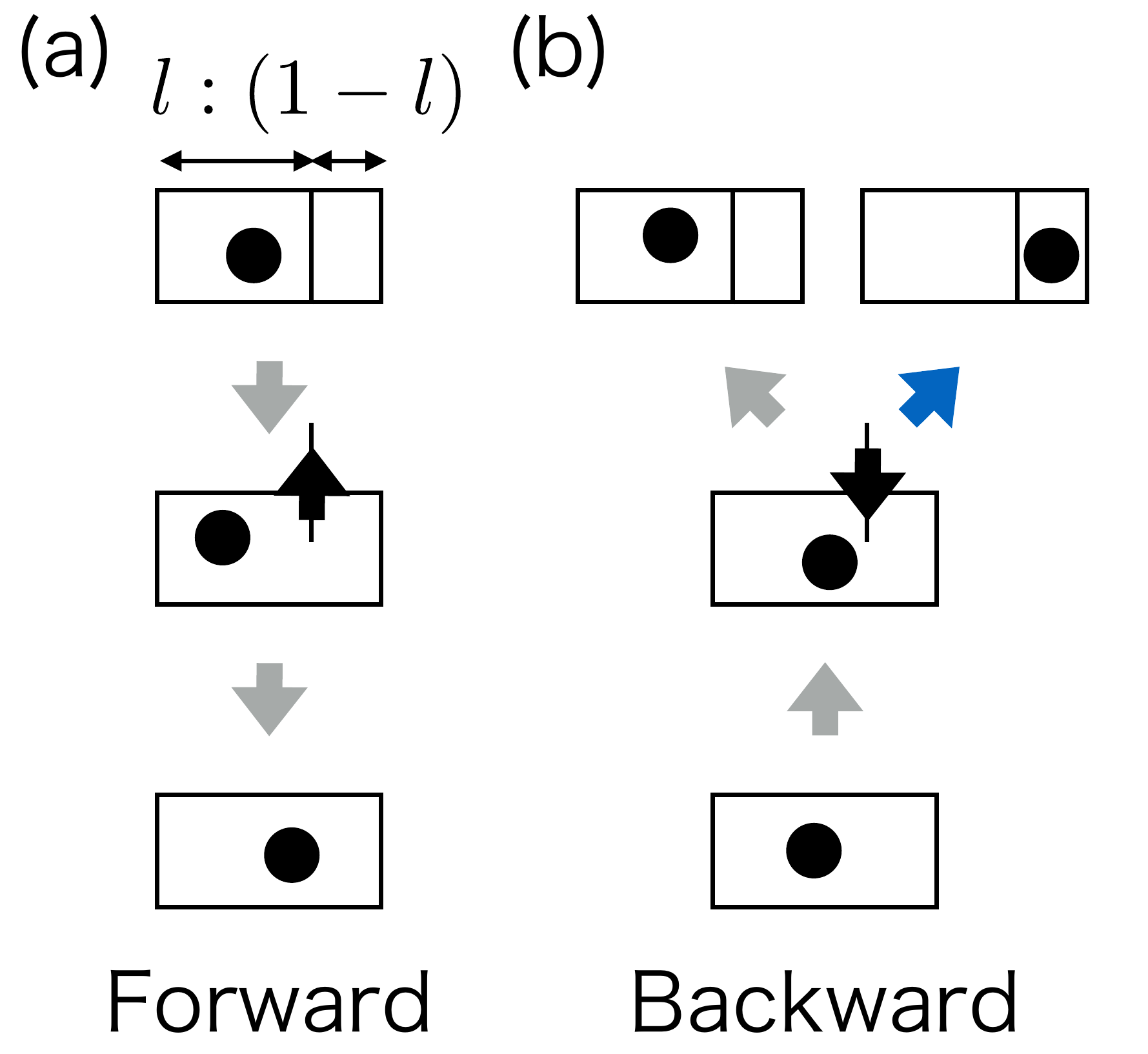}
\caption{
(a) Forward protocol of free expansion.
Initially, a single-particle gas is confined in the left box at temperature $T$.
Then, the wall is removed and the gas expands to the entire box.
(b) Backward protocol of free expansion.
Initially, the particle is in equilibrium with respect to the entire box.
Then, the wall is inserted.
The particle is in the left or right box.
The backward path ending in the right box (indicated by the blue arrow) has no corresponding forward path.
This is a singular path with negatively divergent entropy production.
\label{fig:FE}}
\end{center}
\end{figure}
First of all, let us introduce a concept of {\it absolute irreversibility} in an example of free expansion.
A single-particle gas is initially prepared in the left box (Fig.~\ref{fig:FE}(a)).
Then, the wall is removed and the gas expands to the entire box.
In this process, work is not extracted: $W=0$, whereas the free energy decreases: $\Delta F<0$.
Therefore, the dissipated work is always positive ($W-\Delta F >0$),
and the Jarzynski equality \cite{Jar97Apr, Jar97Nov} breaks down \cite{Gro05Aug} because
\eqn{
	\langle e^{-\beta(W-\Delta F)} \rangle <1.
}
Thus, the Jarzynski equality is not satisfied for free expansion.
Physically, this is because free expansion does not satisfy the assumption of the Jarzynski equality, that is, the initial state is not a global but only a local equilibrium \cite{Jar05}.
Below, we elucidate why a local equilibrium state cannot be the initial state when we derive the conventional nonequilibrium equalitiy.

We consider a set of virtual paths $\{\Gamma_{\rm R}\}$ starting from the right box.
By definition, these paths have vanishing probability for the forward process: $\mP[\Gamma_{\rm R}]=0$.
On the other hand, probability of the paths that end in the right box in the backward process is nonvanishing: $\mP^\dag[\Gamma^\dag_{\rm R}]\neq 0$.
Therefore, we have
\eqn{\label{AI}
	^\exists\Gamma,\ \mP[\Gamma]=0\ \&\ \mP^\dag[\Gamma^\dag]\neq 0.
}
This fact causes the conventional integral fluctuation theorem to break down because
\eqn{
	\langle e^{-\sigma} \rangle
	:&=& \int_{\mP[\Gamma]\neq 0} e^{-\sigma}\mP[\Gamma] \mD\Gamma\nonumber\\
	&=& \int_{\mP[\Gamma]\neq 0} \mP^\dag[\Gamma^\dag] \mD\Gamma
	\nonumber\\
	&=& 1 - \int_{\mP[\Gamma]= 0} \mP^\dag[\Gamma^\dag] \mD\Gamma < 1,
}
where the last inequality follows from Eq.~(\ref{AI}).
On the other hand, in the context of the detailed fluctuation theorem (Eq.~(\ref{CFT})), entropy production is negatively divergent for the paths described by Eq.~(\ref{AI}):
\eqn{
	e^{-\sigma} = \frac{\mP^\dag[\Gamma^\dag_{\rm R}]}{\mP[\Gamma_{\rm R}]} = \infty.
}
We thus conclude that this negatively divergent entropy production is what makes the conventional integral fluctuation equality inapplicable to the process starting from a local equilibrium.

This situation (Eq.~(\ref{AI})) makes a stark contrast to ordinary irreversible processes in which
\eqn{\label{SR}
	^\forall \Gamma,\ \mP[\Gamma]=0\ \Rightarrow\ \mP^\dag[\Gamma^\dag] = 0,
}
In this case, irreversibility is quantified by a finite exponentiated entropy production defined by the detailed fluctuation theorem (\ref{CFT}).
In ordinary irreversible cases, if the backward-path probability is nonzero, then the corresponding forward-path probability is nonzero (Eq.~(\ref{SR})).
Therefore, the paths are stochastically reversible, although they are thermodynamically irreversible.
In contrast, if the condition~(\ref{AI}) holds, there exist paths which are not even stochastically reversible.
Thus, we shall call this process {\it absolutely irreversible}.

The probability ratio $\mP^\dag/\mP$ in the left-hand side of Eq.~(\ref{CFT}) can be interpreted as the transformation function of two probability densities.
In mathematics, probabilities are discussed in measure theory.
Thus, in the next section, we will give a mathematical definition of absolute irreversibility in terms of measure theory and derive nonequilibrium equalities in absolutely irreversible processes.

\subsection{Nonequilibrium Equalities}
\subsubsection{General Formulation}
\begin{figure}
\begin{center}
\includegraphics[width = 1.0\columnwidth]{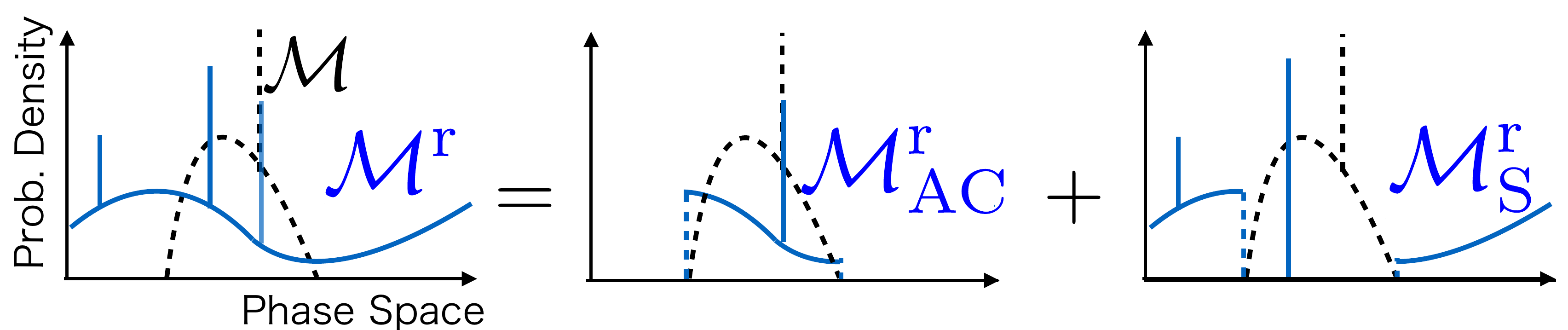}
\caption{
Schematic illustration of the Lebesgue decomposition.
The abscissa represents coordinates of phase space and the ordinate shows the probability density.
Vertical lines represent delta-function-like localization.
Here, $\mM^{\rm r}$ (solid curve) is decomposed into two parts for a given $\mM$ (dashed curve).
In the absolutely continuous part, the probability ratio of $\mM^{\rm r}_{\rm AC}$ to $\mM$ is well-defined.
In the singular part, the ratio of $\mM^{\rm r}_{\rm S}$ to $\mM$ is divergent.}
\label{fig:LD1}
\end{center}
\end{figure}

Let us consider an arbitrary nonequilibrium process without feedback control.
Let $\Gamma(t)$ denote the trajectory of the system in phase space during a time interval $0\le t \le \tau$, and let $\mM[\mD\Gamma]$ and $\mM^{\rm r}[\mD\Gamma]$ denote the path-probability measure in phase space and an arbitrary reference path-probability measure, respectively.
The reference probability is often set to be the probability for the time-reversed dynamics \cite{Jar00, Cro98, Cro99, Cro00, SU10, Sag11, SU12, SU12E, HV10}.
According to Lebesgue's decomposition theorem \cite{Hal74, Bar95},
$\mM^{\rm r}$ can be uniquely decomposed into two parts:
\eqn{\label{decom}
	\mM^{\rm r} = \mM^{\rm r}_{\rm AC} + \mM^{\rm r}_{\rm S},
}
where $\mM^{\rm r}_{\rm AC}$ and $\mM^{\rm r}_{\rm S}$ are absolutely continuous and singular with respect to $\mM$, respectively (Fig.~\ref{fig:LD1}).
The absolute continuity of $\mM^{\rm r}_{\rm AC}$ guarantees that the ratio of $\mM^{\rm r}_{\rm AC}$ to $\mM$ is well-defined.
Physically, it is this ratio that gives entropy production.
On the other hand, $\mM^{\rm r}_{\rm S}$ corresponds to the region  in which the probability defined by $\mM$ vanishes but the probability defined by $\mM^{\rm r}$ remains nonvanishing.
For this part, the ratio of $\mM^{\rm r}_{\rm S}$ to $\mM$ is divergent and we cannot define entropy production through this ratio.
Therefore, $\mM^{\rm r}_{\rm S}$ corresponds to the absolutely irreversible part.
If $\mM^{\rm r}_{\rm S}$ exists, the conventional Jarzynski-type equalities break down \cite{Sun05, LG05, Gro05Aug, Jar05, Gro05Sep, SU12, FWU13}.

In nonequilibrium equalities, we evaluate ensemble averages of path functionals both in the original probability measure and in the reference one.
Entropy production is a key quantity which connects the reference average with the original one.
The transformation between the two probability measures is mathematically formulated by the Radon-Nikodym theorem.
Because $\mM^{\rm r}_{\rm AC}$ is absolutely continuous with respect to $\mM$, we may apply the Radon-Nikodym theorem to obtain
\eqn{
	\mM^{\rm r}_{\rm AC}[\mD\Gamma] =
	\left. \frac{\mD\mM^{\rm r}_{\rm AC}}{\mD\mM} \right|_{\Gamma} \mM[\mD\Gamma],
}
where
$\left. {\mD\mM^{\rm r}_{\rm AC}}/{\mD\mM} \right|_{\Gamma}$
is the Radon-Nykodym derivative which is an integrable function with respect to $\mM$; it is nothing but the ratio of two probabilities in the present context.
Let us formally define the entropy production as
\eqn{\label{entropy}
	\sigma = -\ln\left.\frac{\mD\mM^{\rm r}_{\rm AC}}{\mD\mM}\right|_{\Gamma}.
}
If $\mM$ and $\mM^{\rm r}_{\rm AC}$ can be written by probability densities $\mM[\mD\Gamma]=\mP[\Gamma]\mD\Gamma$ and $\mM^{\rm r}_{\rm AC}[\mD\Gamma]=\mP^{\rm r}[\Gamma]\mD\Gamma$, Eq.~(\ref{entropy}) can be rewritten as
$
	\sigma = -\ln {\mP^{\rm r}[\Gamma]}/{\mP[\Gamma]},
$
which is the standard definition of entropy production.
A physical interpretation of $\sigma$ will be discussed later.
Let $\mF$ denote an arbitrary functional of a path and let $\langle \cdots \rangle$, $\langle \cdots \rangle^{\rm r}$ and $\langle \cdots \rangle^{\rm r}_I\ (I={\rm AC}, {\rm S})$ denote the averages over $\mM$, $\mM^{\rm r}$ and $\mM^{\rm r}_I$, respectively.
Then, we can evaluate the average of the absolutely continuous part by entropy production:
\eqn{
	\langle \mF \rangle^{\rm r}_{\rm AC} &=& \int \mF[\Gamma]\mM^{\rm r}_{\rm AC}[\mD\Gamma]\nonumber\\
	&=&\int \mF[\Gamma]\left. \frac{\mD\mM^{\rm r}_{\rm AC}}{\mD\mM} \right|_{\Gamma} \mM[\mD\Gamma]\nonumber\\
	&=& \langle \mF e^{-\sigma} \rangle.
}
In accordance with the decomposition in Eq.~(\ref{decom}),
$
	\langle \mF \rangle_{\rm AC}^{\rm r}
	=\langle \mF \rangle^{\rm r} - \langle \mF \rangle_{\rm S}^{\rm r}
$
holds. Thus we obtain the following modified integral fluctuation theorem:
\eqn{\label{FJar}
	\langle \mF e^{-\sigma} \rangle
	=\langle \mF \rangle^{\rm r} - \langle \mF \rangle_{\rm S}^{\rm r}.
}
This equality may be regarded as a generalization of the master integral fluctuation theorem in Refs.~\cite{Cro00, Sei12}.
If we set $\mF=1$, Eq.~(\ref{FJar}) reduces to
\eqn{\label{Jar}
	\langle e^{-\sigma} \rangle = 1-\lambda_{\rm S},
}
where $\lambda_{\rm S} = \int \mM^{\rm r}_{\rm S}[\mD\Gamma(t)]$ is the probability of the singular part, or absolutely irreversible part.
This absolute irreversibility arises from two kinds of divergent entropy.
One is the case in which the probability ratio is formally written as $\mP^{\rm r}/\mP=\mP^{\rm r}/0$, {\it i.e.}, the forward probability vanishes but the reference probability does not as in free expansion.
The other is the case in which $\mP^{\rm r}/\mP=\delta(0)/\mP$, {\it i.e.}, the reference probability has delta-function-like singularity but the forward probability does not as in a setting with localized traps.
(In this case, $\mM[D]$ is zero, whereas $\mM^{\rm r}[D]$ is not, where $D$ is an infinitesimal region around the center of the delta function.)
Because $\lambda_{\rm S}$ is the total probability of these two cases, we can calculate $\lambda_{\rm S}$ by summing up the probabilities of those reference paths that have vanishing probability for the corresponding original paths and that of the localized reference paths.
To the best of our knowledge, no research has been conducted on the localized case.
However, the existence of this part renders the conventional fluctuation theorem inapplicable.
The Jarzynski equality $\langle e^{-\sigma}\rangle=1$ is reproduced only when $\lambda_{\rm S} = 0$.

Using Jensen's inequality $\langle e^{-\sigma} \rangle \ge e^{-\langle \sigma \rangle}$, we have
\eqn{\label{Jarin}
	\langle \sigma \rangle \ge - \ln (1-\lambda_{\rm S}).
}
This inequality indicates that the second law of thermodynamics holds even in absolutely irreversible cases because the right-hand side of Eq.~(\ref{Jarin}) is equal to or greater than zero.
Moreover, when the process involves the absolutely irreversible paths as in free expansion, {\it i.e.}, when $\lambda_{\rm S}>0$, the inequality~(\ref{Jarin}) imposes a stronger restriction on the average entropy production than the second law, that is, the average entropy production must be positive in this case.

\begin{figure}
\begin{center}
\includegraphics[width = 1.0\columnwidth]{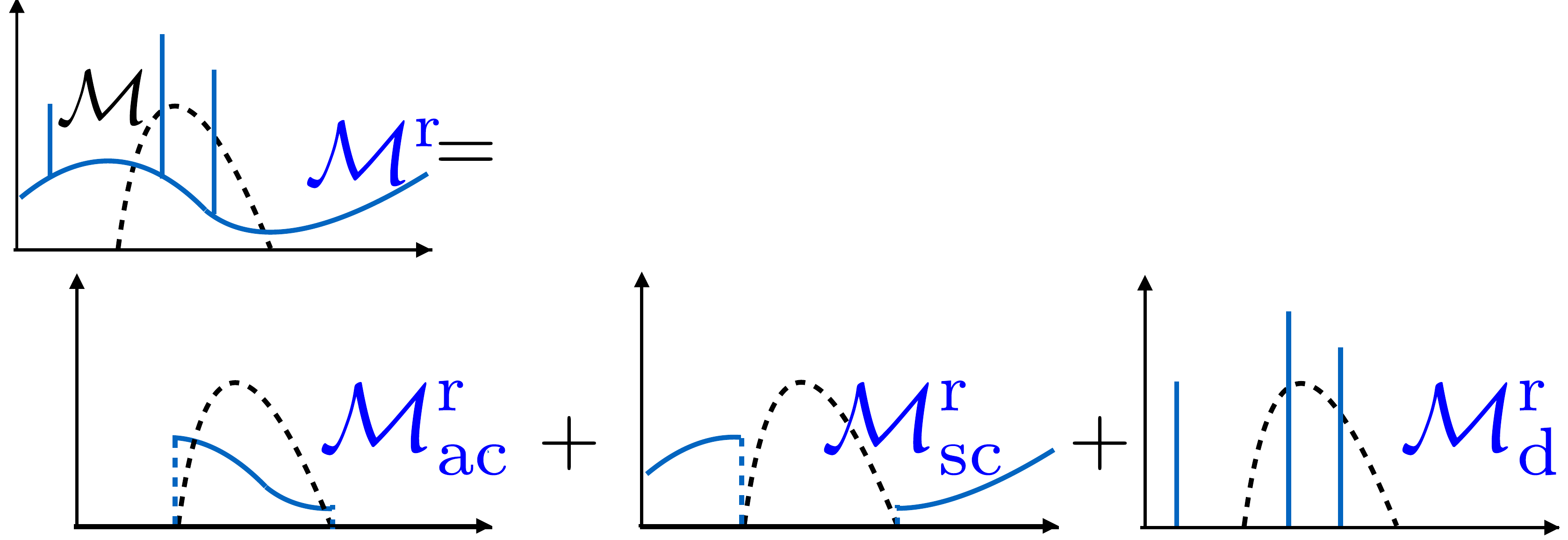}
\caption{
Schematic illustration of the stronger version of the Lebesgue decomposition.
Here, $\mM^{\rm r}$ (solid curve) is decomposed into three parts for a given $\mM$ (dashed curve).
In the absolutely continuous part (lower left), the probability ratio of $\mM^{\rm r}_{\rm ac}$ to $\mM$ is well-defined.
In the singular continuous part (lower middle), the ratio of $\mM^{\rm r}_{\rm sc}$ to $\mM$ is divergent because the probability density of $\mM$ vanishes.
In the discrete part (lower right), the ratio of $\mM^{\rm r}_{\rm d}$ to $\mM$ is divergent because the probability density of $\mM^{\rm d}$ diverges whereas $\mM$ is finite.
Note that $\mM$ involves no delta-function-like localization, which is the assumption for the stronger Lebesgue decomposition and needed to guarantee the uniqueness of the decomposition.}
\label{fig:LD2}
\end{center}
\end{figure}

When $\mM$ can be written in terms of the probability density, a stronger version of Lebesgue's decomposition holds.
Now, $\mM^{\rm r}$ can be uniquely decomposed into three parts:
\eqn{\label{3decom}
	\mM^{\rm r} = \mM^{\rm r}_{{\rm ac}}
		+ \mM^{\rm r}_{{\rm sc}}
		+\mM^{\rm r}_{{\rm d}},
}
where $\mM^{\rm r}_{{\rm ac}}$ is absolutely continuous with respect to $\mM$,
$\mM^{\rm r}_{{\rm sc}}$ is the singular continuous part corresponding to the vanishing forward-path probability, and
$\mM^{\rm r}_{{\rm d}}$ is the discrete part corresponding to delta-function like localization (Fig.~\ref{fig:LD2}).
The second term represents the effect of expansion and the third term represents the effect of trapping.
In accordance with Eq.~(\ref{3decom}), we obtain
\eqn{\label{3FSU}
	\langle \mF e^{-\sigma} \rangle &=&
		\langle \mF \rangle^{\rm r} -
		\langle \mF \rangle^{\rm r}_{\rm sc} -
		\langle \mF \rangle^{\rm r}_{\rm d},\\
	\langle e^{-\sigma} \rangle &=& 1-\lambda_{\rm sc}-\lambda_{\rm d},\label{3SU}
}
where $\langle \cdots \rangle^{\rm r}_i\ (i={\rm sc}, {\rm d})$ denotes the averages over $\mM^{\rm r}_i$.

\subsubsection{Physical Implications}
The proof of nonequilibrium equalities (\ref{FJar}), (\ref{Jar}), (\ref{3FSU}) and (\ref{3SU}) in the previous section can be made on a very general ground regardless of the dynamics of the system.
In particular, the proof can be applied to Langevin systems and Hamiltonian systems.
Physics enters the problem in the choice of the reference probability measure.
Once the reference probability measure is properly chosen, the formal entropy production discussed above becomes the corresponding physical entropy production, which is measurable in experiments.
This fact is widely known in the absence of absolute irreversibility \cite{Sei12} and explicitly shown in Appendix A.
Here, we will briefly review the main results.

By comparing the original dynamics with the time-reversed one, we can quantify the asymmetry of thermodynamic processes with respect to time reversal.
Here, we set the reference dynamics to the time-revered dynamics under the time-reversed protocol.

If we assume that the initial probability of the original process is a local canonical distribution with inverse temperature $\beta$, and set the initial probability of the time-reversed path to another local canonical distribution with the same inverse temperature $\beta$, then the entropy production reduces to
\eqn{\label{DW}
	\sigma = \beta(W-\Delta F),
}
where $W$ is the work performed on the system and $\Delta F$ is the free-energy difference of the system \cite{Sei12}.
Thus, the integral fluctuation theorem is
\eqn{
	\langle e^{-\beta(W-\Delta F)}\rangle = 1-\lambda_{\rm S},
}
which reduces to the Jarzynski equality \cite{Jar97Apr, Jar97Nov} only when $\lambda_{\rm S}=0$.
In this case, the value of $\lambda_{\rm S}$ can be experimentally determined through the measurement of the time-reversed process; $\lambda_{\rm S}$ is the total probability of localized backward paths and backward paths for which the corresponding forward paths are absent.
In special cases, we can analytically calculate $\lambda_{\rm S}$ as shown in Sec. II C 2.
The corresponding inequality
\eqn{
	W^{\rm ext} \le -\Delta F + \ln(1-\lambda_{\rm S})
}
demonstrates that the extractable work $W^{\rm ext}(:= -W)$ is diminished by the effect of absolutely irreversible processes since $\ln(1-\lambda_{\rm S}) \le 0$.

When we set the initial probability distribution of the time-reversed process to the final distribution of the original one, we obtain
\eqn{
	\langle e^{-\Delta s^{\rm tot}}\rangle = 1-\lambda_{\rm S},
}
where $\Delta s^{\rm tot}$ is the sum of the Shannon entropy production of the system and the entropy production of the bath accompanying the absorption of heat \cite{Sei12}.

Finally, we note that other choices of the reference probability lead to fluctuation theorems on other types of entropy production such as housekeeping entropy production and excess entropy production \cite{Sei12}.

\section{Examples of Absolutely Irreversible Processes}
When the initial probability distribution has no
delta-function-like localization, the stronger version of the Lebesgue decomposition holds, and the nonequilibrium equality in absolutely irreversible processes is given by
\eqn{\label{main}
	\langle e^{-\sigma} \rangle = 1-\lambda_{\rm sc}-\lambda_{\rm d}.
}
The term $\lambda_{\rm sc}$ is the singular continuous part and represents the effect of free expansion.
We can calculate $\lambda_{\rm sc}$ by summing up the probabilities of those time-reversed paths whose corresponding forward paths vanish.
The term $\lambda_{\rm d}$ is the discrete part, which represents the effect of trapping.
We can calculate $\lambda_{\rm d}$ by summing up the probability of spatially localized time-reversed paths.
We will demonstrate the validity of Eq.~(\ref{main}) in three examples below.

\subsection{Free Expansion}
We first discuss free expansion (see Fig.~\ref{fig:FE} (a)).
In this example, we assume that the single-particle gas is initially in the local equilibrium in the left box, where the volume ratio between the left and right boxes is taken to be $l:1-l$.
Therefore, we choose $\sigma$ as in Eq.~(\ref{DW}):
\eqn{\label{C1res}
	\langle e^{-\beta(W-\Delta F)} \rangle = 1-\lambda_{\rm sc} -\lambda_{\rm d}.
}
The wall is removed and the free energy of the gas decreases by
$
	\Delta F = k_BT\ln l(<0).
$
In this process, work is neither extracted nor performed: $W=0$.
Therefore, the exponentiated average is
\eqn{\label{C1ave}
	\langle  e^{-\beta(W-\Delta F)} \rangle = l.
}

We next consider the backward process (see Fig.~\ref{fig:FE} (b)).
In the backward process, the initial state of the single-particle gas is in equilibrium over the entire box.
Then, a wall is inserted at the same position as the forward process.
The particle is in either left or right box.
The paths ending in the right box have no corresponding paths in the forward process.
Therefore, these paths are singular continuous.
The probability of these paths is proportional to the volume fraction of the right box:
\eqn{\label{C1sc}
	\lambda_{\rm sc} = 1-l.
}
Since there is no single path with finite probability (such a path would contribute to the discrete singularity),
\eqn{\label{C1d}
	\lambda_{\rm d} = 0.
}
We can easily confirm that Eqs.~(\ref{C1ave}), (\ref{C1sc}) and (\ref{C1d}) are consistent with Eq.~(\ref{C1res}).

In fact, Eq.~(\ref{C1ave}) can be derived by conventional methods as well and this is true for the following two examples.
In Appendix B, we compare our method with the conventional ones and discuss advantages of our method.
In our method, the meaning of the obtained nonequilibrium equality is much clearer and the derivation is simpler because our method directly deals with the process of interest.

\subsection{Process Starting from a Local Equilibrium}
\begin{figure}
\begin{center}
\includegraphics[width = 0.5\textwidth]{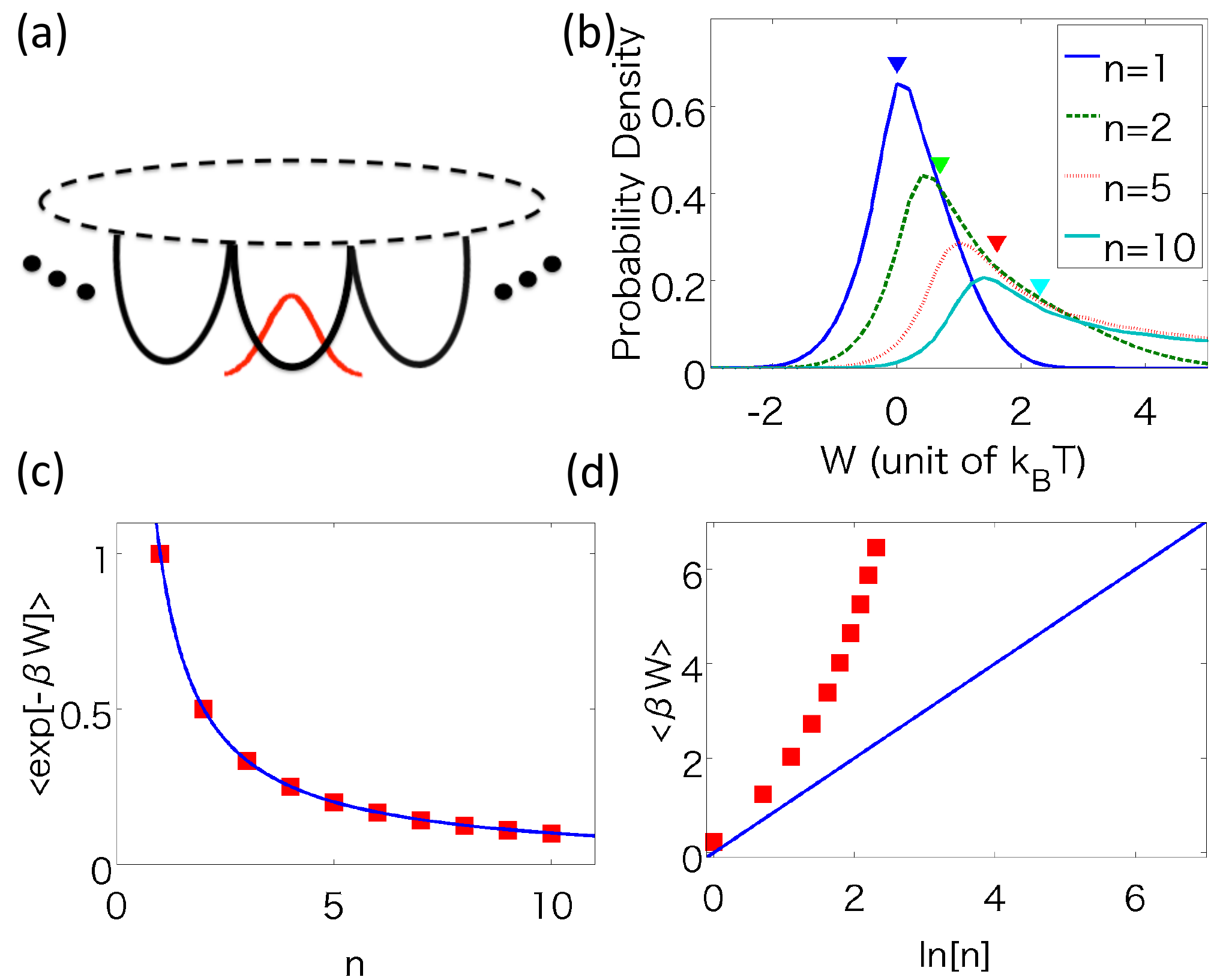}
\caption{\label{sim}
(a) Schematic illustration of an overdamped Langevin system consisting of $n$ harmonic potential wells on a one-dimensional ring.
(b) Probability density of work performed on the system for several values of $n$. The triangles indicate the points of $W=k_BT\ln n$.
(c) Averaged values of $\langle \exp[{-\beta W}] \rangle$ at each $n$. Superimposed is a fitted $1/n$ curve.
(d) Averaged values of $\langle \beta W \rangle$. The line represents the minimum dissipation given by $k_BT\ln n$.
The parameters are chosen as follows: diffusion constant $D=10^{-13}{\rm m^2/s}$; temperature $T=300{\rm K}$; duration of the process $\tau = 10{\rm sec}$; half width of a single potential $a=10^{-6}{\rm m}$; the initial stiffness of potential $K$ is chosen so as to satisfy $Ka^2/2=5k_BT$. The nonequilibrium process is repeated $10^6$ times for each $n$. 
}
\end{center}
\end{figure}

Next, let us again consider the case of $\lambda_{\rm sc} \neq 0$, {\it i.e.}, the case in which the singular continuous part exists in a more complicated system.
We perform numerical simulations for an overdamped Langevin system confined to a one-dimensional ring, where the potential consists of $n$ identical harmonic potential wells with stiffness (namely, the spring constant) $k(t)$ (see Fig.~\ref{sim}(a)).
The initial distribution is set to be the local equilibrium distribution in a given well and vanishes elsewhere.
We set the reference initial probability to the canonical distribution corresponding to the final potential and set the reference dynamics to the time reversal dynamics.
We study a nonequilibrium process during a time interval $\tau$, in which the stiffness of potentials is decreased from $k=K$ to $0$ at a constant rate between $t=0$ and $\tau/2$, and then increased from $k=0$ to $n^2K$ at a constant rate between $t=\tau/2$ and $\tau$.
Because a backward path terminates in a certain well with probability $1/n$ due to the symmetry of the potential and the initial backward state, the probability that the backward path does not have the corresponding forward path is $\lambda_{\rm sc}=(n-1)/n$.
Then the nonequilibrium integral equality~(\ref{main}) reduces to
\eqn{\label{sc}
	\langle e^{-\beta(W-\Delta F)} \rangle = \frac{1}{n},
}
and the corresponding second law is given by
\eqn{\label{ineq}
	\langle W \rangle \ge \Delta F +  k_BT \ln n.
}
If we assume $K$ is sufficiently large, 
$\Delta F$ is zero in this process.
Figure~\ref{sim}(b) shows the distribution of work at different $n$ obtained by numerical simulations.
The value of $\langle e^{-\beta W} \rangle$ is confirmed to be $1/n$ as demonstrated in Fig.~\ref{sim}(c).
It is also verified that the averaged value of dissipation is larger than the minima predicted by the inequality (\ref{ineq}) (Fig.~\ref{sim}(d)).
When the initial probability distribution is localized, there is diffusion to the entire phase space; then entropy production tends to be positive.
Note that this process can be regarded as an information erasure process of the symmetric $n$-bit memory.

\subsection{System with a Trap}
\begin{figure}[t]
\includegraphics[width = 0.48\textwidth]{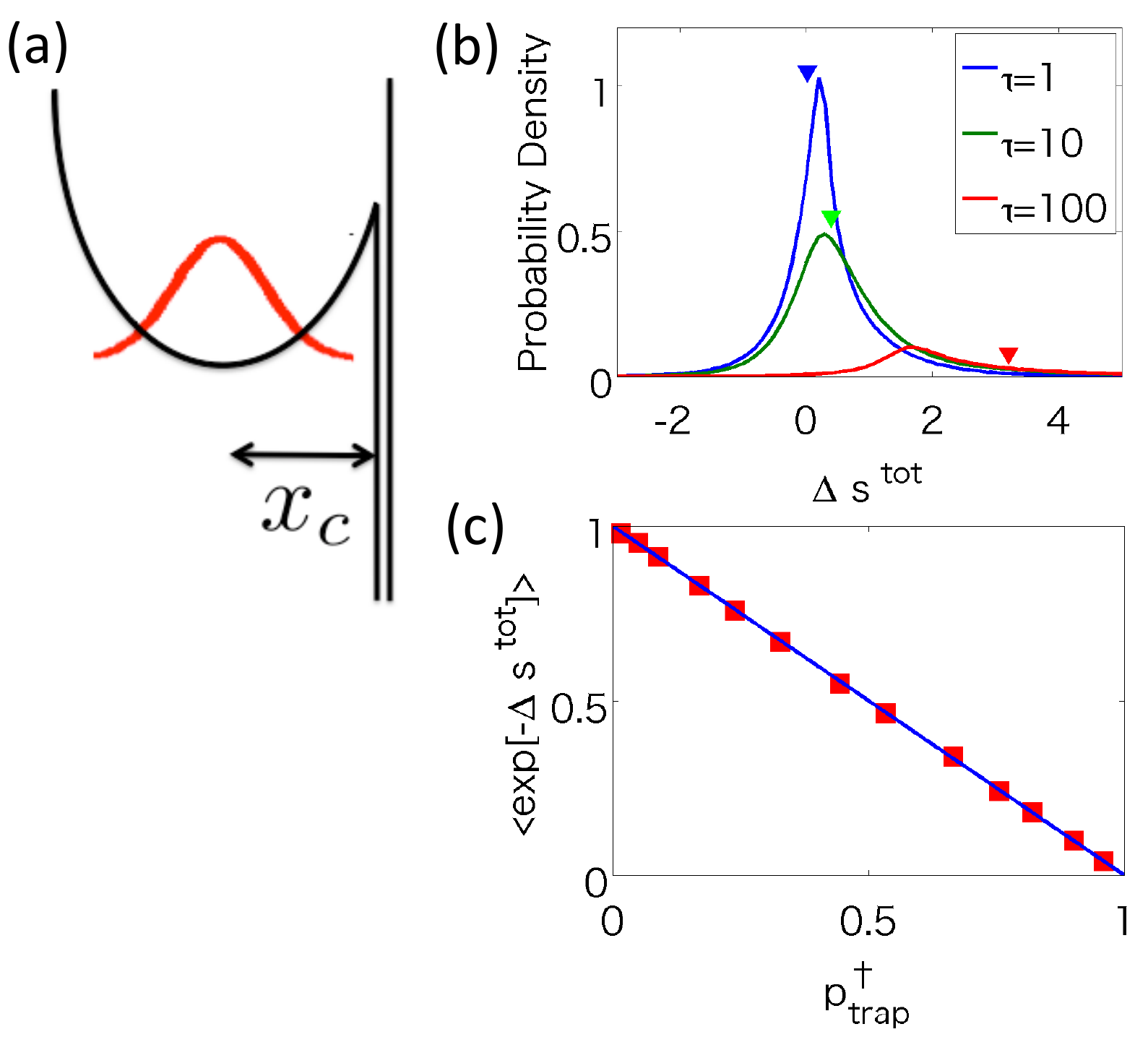}
\caption{\label{sim2}
(a) An overdamped Langevin system with a trap.
(b) Probability density of the total entropy production for several values of $\tau$. The triangles indicate the points where $\Delta s^{\rm tot}= -\ln(1-p_{\rm trap}^\dag)$.
(c) Averaged values of $\langle\exp[-\Delta s^{\rm tot}]\rangle$ versus the trapping probability $p_{\rm trap}^\dag$.
Superimposed is a fitted $1-p_{\rm trap}^\dag$ line.
The parameters $D$ and $T$ are the same as in Fig.~1.
The distance between the trap and the center of the potential is $x_c = 10^{-6}{\rm m}$.
The initial stiffness $K$ is set so as to satisfy $Ka^2/2 = 10k_BT$.
The duration of the process $\tau$ is varied between $1 {\rm sec}$ and $100 {\rm sec}$. The process is repeated $10^6$ times for each $\tau$.
}
\end{figure}

Finally, we consider the case of $\lambda_{\rm d}\neq 0$, {\it i.e.}, the case in which the discrete singularity exists.
We perform numerical simulations for a one-dimensional system in which a single particle is confined in a single harmonic potential with stiffness $k(t)$.
We assume that there is a trapping point in the system and that the distance between the point and the center of the harmonic potential is $x_c$ (see Fig.~\ref{sim2}(a)).
If the particle reaches the trapping point, it will be trapped with unit probability.
The initial distribution is the equilibrium distribution of the harmonic potential.
The stiffness of the potential is decreased from $k=K$ to $0$ at a constant rate between $t=0$ and $\tau/2$, and then increased from $0$ to $K$ at a constant rate between $t=\tau/2$ to $\tau$.
To derive a nonequilibrium equality, we set the reference initial probability distribution to the final probability distribution of the original process, and set the reference dynamics to the time reversal dynamics.
Let $p_{\rm trap}$ and $p_{\rm trap}^\dag$ be the trapping probabilities of the final states of the original and time-reversed processes, respectively.
Because the trapped particle remains trapped in the entire time-reversed process, the probability of this single backward path is $p_{\rm trap}(\neq 0)$, and thus the discrete probability is $\lambda_{\rm d}=p_{\rm trap}$.
Moreover, since the paths that fall into the trap in the time-reversed process have no corresponding paths in the original process, they are singular continuous.
Therefore, the singular continuous probability is $\lambda_{\rm sc}=p_{\rm trap}^\dag - p_{\rm trap}$.
Thus, Eq.~(\ref{main}) reduces to
\eqn{\label{d}
	\langle e^{-\Delta s^{\rm tot}} \rangle = 1-p_{\rm trap}^\dag,
}
which is consistent with numerical simulations as shown in Fig.~\ref{sim2}(c).
The corresponding inequality
\eqn{
	\langle \Delta s^{\rm tot} \rangle \ge -\ln (1-p_{\rm trap}^\dag)
}
is automatically satisfied because the left-hand side is positively divergent due to those paths that fall into the trap in the original process with positively divergent entropy.

\section{Nonequilibrium Equalities in Absolutely Irreversible Processes under Measurements and Feedback Control}
\subsection{Main Equalities}
Let us now consider a nonequilibrium process with feedback control.
Under feedback control, it is important to consider the singular part because high-precision measurements such as error-free measurements necessarily localize the probability distribution and this localization promotes subsequent free expansion. 

Let $\Lambda(t)$ denote a path in space of outcomes.
The feedback control protocol, which determines the time evolution of $\Gamma(t)$, depends on $\{\Lambda(s)\}_{s=0}^t$.
In other words, the control parameters are regulated based on the history of the outcome $\{\Lambda(s)\}_{s=0}^t$,
and the transition probability from $\Gamma(t)$ to $\Gamma(t+dt)$ is fixed by these parameters (and by the history of the system $\{\Gamma(s)\}_{s=0}^t$).
On the other hand, the time evolution of the outcome $\Lambda(t)$ is determined by measurements which depend on $\{\Gamma(s)\}_{s=0}^t$,
that is, we obtain an outcome $\Lambda(t)$ based on the history of the system $\{\Gamma(s)\}_{s=0}^t$.
These measurements and control protocol determine the entire time evolution of $\Gamma(t)$ and $\Lambda(t)$.
Therefore, the entire paths $\Gamma:=\{\Gamma(t)\}_{t=0}^\tau$ and $\Lambda:=\{\Lambda(t)\}_{t=0}^\tau$ become correlated due to the feedback control and measurements.
For a given entire history of the outcome $\Lambda$, let $\mM_{|\Lambda}$ denote the conditional probability measure for paths whose measurement outcome is given by $\Lambda$,
and let us choose an arbitrary reference probability measure $\mM^{\rm r}_{|\Lambda}$, which can also be decomposed into two parts with respect to $\mM_{|\Lambda}$, {\it i.e.}, 
$
	\mM^{\rm r}_{|\Lambda}
	=\mM^{\rm r}_{{\rm AC}|\Lambda}+\mM^{\rm r}_{{\rm S}|\Lambda}.
$
Once $\Lambda$ is given, the control protocol is also fixed.
Therefore, we obtain
\eqn{
	\langle \mF e^{-R_{|\Lambda}} \rangle_{|\Lambda}
	= \langle \mF \rangle^{\rm r}_{|\Lambda}
	- \langle \mF \rangle^{\rm r}_{{\rm S}|\Lambda}, 
}
where
\eqn{
	R_{|\Lambda} =
	-\ln \left.
		\frac{\mD\mM^{\rm r}_{{\rm AC}|\Lambda}}{\mD\mM_{|\Lambda}}
	\right|_{\Gamma}.
}
Averaging this with respect to the path of the outcome $\Lambda$, we obtain
\eqn{\label{preSU}
	\langle \mF e^{-R_{|\Lambda}} \rangle
	= \langle \mF \rangle^{\rm r}
	- \langle \mF \rangle^{\rm r}_{\rm S} .
}
The entropy production of the system should be defined by the ratio of the reference transition probability to the original one under a fixed $\Lambda$ or fixed protocol:
\eqn{
	\sigma &=& -\ln \left.
		\frac{\mD\mM^{\rm r}_{{\rm AC}|\Lambda}}{\mD\mM^{\rm trans}_{|\Lambda}}
	\right|_{\Gamma},
}
where $\mM^{\rm trans}_{|\Lambda}$ is the transition probability of $\Gamma$ under a given $\Lambda$.
The reference transition probability is nothing but the conditional probability because there are no measurements.
On the other hand, the original transition probability $\mM^{\rm trans}_{|\Lambda}$ is different from the original conditional probability $\mM_{|\Lambda}$ \cite{SU12E}.
Then $R_{|\Lambda}$ can be decomposed into two parts:
\eqn{
	R_{|\Lambda} = -\ln \left.
		\frac{\mD\mM^{\rm r}_{{\rm AC}|\Lambda}}{\mD\mM^{\rm trans}_{|\Lambda}}
	\right|_{\Gamma}
	-\ln \left.
		\frac{\mD\mM^{\rm trans}_{|\Lambda}}{\mD\mM_{|\Lambda}}
	\right|_{\Gamma}
		= \sigma + I,\ \ \ 
}
where
$
	I := -\ln{\mD\mM^{\rm trans}_{|\Lambda}}/{\mD\mM_{|\Lambda}}|_{\Gamma}. 
$
The transition probability $\mM^{\rm trans}_{|\Lambda}$ quantifies the correlation originating from feedback control, while the conditional probability $\mM_{|\Lambda(t)}$ contains the correlation originating from both measurements and feedback control.
Because $I$ quantifies the difference between these two probability, $I$ may be regarded as the correlation originating from measurements, or mutual information.
If $\mM^{\rm trans}_{|\Lambda}$ and $\mM_{|\Lambda}$ can be expressed in terms of probability densities, we can show directly that this definition reduces to the standard definition of mutual information \cite{SU12E}:
$
	I = -\ln{\mP[\Lambda]}/{\mP[\Lambda|\Gamma]}.
$
Equation (\ref{preSU}) can be rewritten as
\eqn{\label{FSU}
	\langle \mF e^{-\sigma-I} \rangle
	= \langle \mF \rangle^{\rm r}
	- \langle \mF \rangle^{\rm r}_{\rm S}.
}
If we set $\mF=1$, Eq.(\ref{FSU}) reduces to
\eqn{\label{SU}
	\langle e^{-\sigma - I} \rangle = 1 - \overline{\lambda}_{\rm S},
}
where $\overline{\lambda}_{\rm S}$ is the average over $\Lambda$ of the outcome-conditioned singular-part probabilities. 
In particular, if the measurement is performed only once and if there are no singular parts in the reference probability, the equality reproduces the original equality obtained in Refs.~\cite{SU10, Sag11}.
The corresponding inequality is
\eqn{\label{IT2}
	\langle \sigma \rangle
	\ge
	-\langle I \rangle - \ln(1-\overline{\lambda}_{\rm S}).
}
Thus, the lower bound of entropy production is determined not only by the mutual information but also by the term arising from absolute irreversibility.
In particular, if $\langle I \rangle > - \ln(1-\overline{\lambda}_{\rm S})$, {\it i.e.}, if the right-hand side of Eq.~(\ref{IT2}) is negative, the averaged entropy production can be negative.

When the initial state satisfies the assumption of the stronger version of the Lebesgue decomposition, we obtain
\eqn{
	\langle e^{-\sigma -I} \rangle
	&=&1-\overline\lambda_{\rm sc} -\overline \lambda_{\rm d},\\
	\langle\sigma\rangle&\ge&
	- \langle I\rangle -\ln(1-\overline\lambda_{\rm sc} -\overline \lambda_{\rm d}),
}
where $\overline\lambda_{\rm sc}$ and $\overline\lambda_{\rm d}$ are the averaged singular continuous and discrete probabilities, respectively.

In a manner similar to the case without feedback control, a proper choice of the reference probability leads to the corresponding fluctuation theorem.
If the initial state is in a local equilibrium, Eq.~(\ref{SU}) reduces to
\eqn{\label{SUinAIP}
	\langle e^{-\beta(W-\Delta F)-I}\rangle = 1-\overline\lambda_{\rm S}.
}
This can be seen if we set the reference dynamics to the time-reversed one and the initial distribution of the time-reversed process to a local canonical one.
When $\overline\lambda_{\rm S}=0$, Eq.~(\ref{SUinAIP}) reduces to the generalized Jarzynski equality under feedback control derived in Refs. \cite{SU10, Sag11, SU12E, HV10}.
On the other hand, if we set the reference dynamics to the time-reversed one and the initial distribution of the time-reversed process to the final distribution of the original process, we obtain
\eqn{
	\langle e^{-\Delta s^{\rm tot}-I}\rangle = 1-\overline\lambda_{\rm S}.
}
In these cases, the value of $\overline\lambda_{\rm S}$ can be calculated via the time-reversed process in a manner similar to the case without feedback control.

\subsection{Examples}
\subsubsection{Measurement and Trivial Feedback Control}
\begin{figure}
\begin{center}
\includegraphics[width = 1.0\columnwidth]{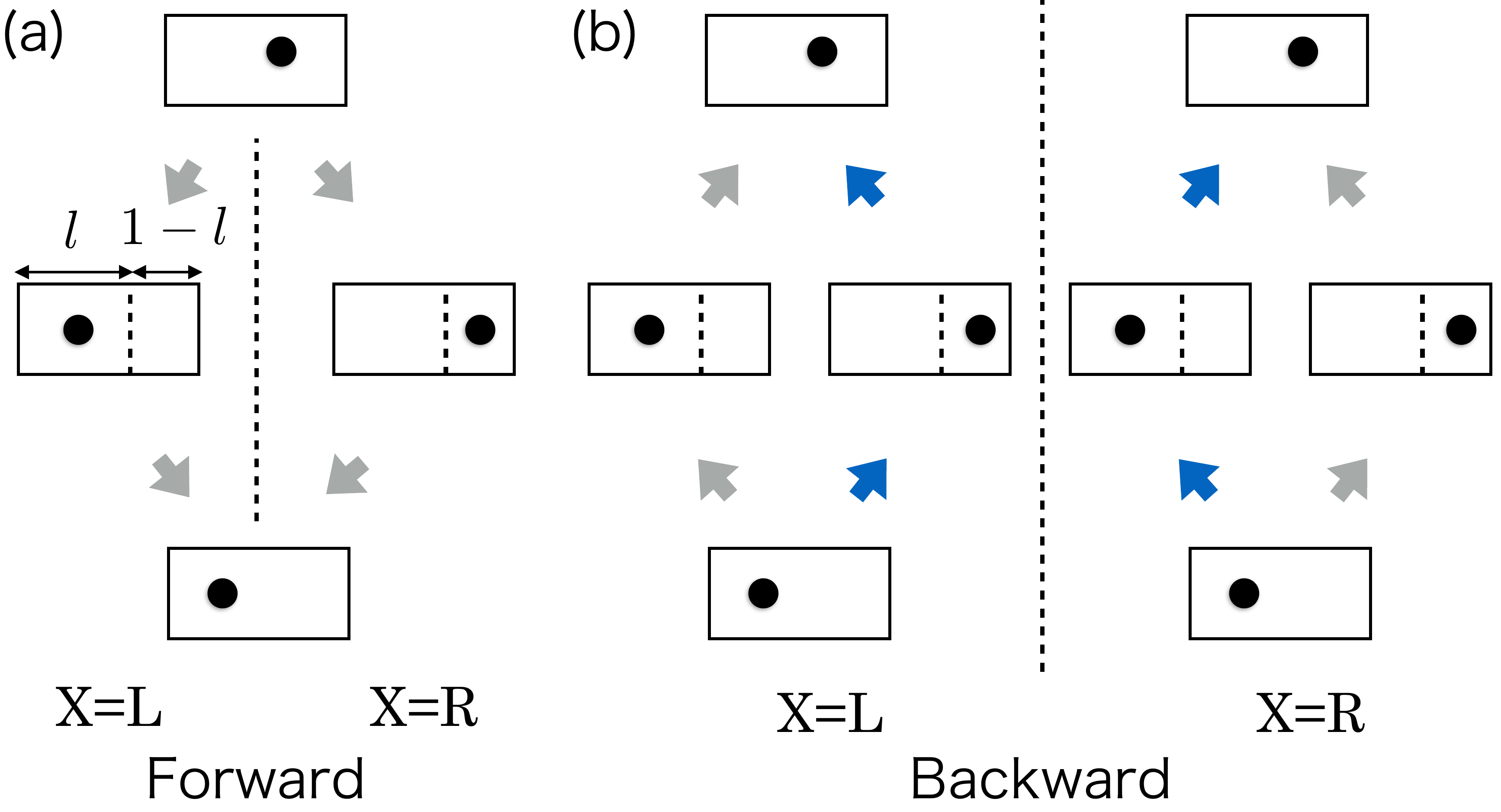}
\caption{
(a) A measurement and the subsequent trivial feedback control.
Initially, a particle is in equilibrium with the entire box.
Then, we perform an error-free position-measurement to determine whether the position $X$ is the left $L$ or right $R$.
After this measurement, we do nothing.
(b) The time-reversed protocol.
The particle is in equilibrium and we do nothing.
At the time of the measurement, the particle is probabilistically in either the left or right side.
For the outcome $X=L$, the case in which the particle is in the right side is singular (blue arrows), and similarly for the outcome $X=R$.
\label{fig:MwoFC}
}
\end{center}
\end{figure}

Let us consider a measurement and the subsequent trivial feedback control to verify Eq.~(\ref{SUinAIP}).
At first, a particle is in the global equilibrium state of the box (Fig.~\ref{fig:MwoFC}(a)).
Then, we perform an instantaneous error-free measurement to determine the position $X$.
We obtain an outcome $X=L$ when the particle is in the left side, or the length from the left-end wall is shorter than the length of the whole box multiplied by $l$.
On the other hand, we obtain an outcome $X=R$ when the particle is in the right side.
In both cases, we do nothing after the measurement.
In this process, work is not extracted: $W=0$, and free energy does not changes: $\Delta F=0$.
By the measurement, when $X=L$, we obtain mutual information $I=-\ln p(L)$, where $p(L)$ is the probability to obtain an outcome $X=L$.
When $X=R$, we obtain $I=-\ln p(R)$.
Therefore, the exponentiated average is
\eqn{\label{SUwoFC}
	\langle e^{-\beta(W-\Delta F)-I}\rangle
	&=& p(L)\cdot e^{\ln p(L)}
	+p(R)\cdot e^{\ln p(R)}\nonumber\\
	&=& p(L)^2 + p(R)^2\nonumber\\
	&=& l^2 + (1-l)^2.
}

Next, let us consider the time-reversed process to calculate the singular probability.
In the time-reversed process, we do nothing for both outcomes $L$ and $R$, because we do nothing in the forward process.
For the time-reversed process of the outcome $X=L$ (the left half of Fig.~\ref{fig:MwoFC}(b)), the particle is in the right side at the time of the measurement with probability $1-l$.
This is the singular event, which never happens in the forward process because of the error-free property of the measurement.
Therefore, we obtain
\eqn{
	\lambda_{{\rm S}|X=L} = 1-l.
}
In a similar manner, we obtain
\eqn{
	\lambda_{{\rm S}|X=R} = l.
}
Thus, the averaged singular probability is calculated as
\eqn{
	\overline\lambda_{\rm S} 
	&=& p(L)\lambda_{{\rm S}|X=L}+p(R)\lambda_{{\rm S}|X=R}\nonumber\\
	&=& 2l(1-l).
}
By a simple calculation, we can confirm that this value is consistent with Eqs.~(\ref{SUinAIP}) and (\ref{SUwoFC}).

After the measurement, the particle freely expands to the entire box because we do no feedback control, and we lose useful information obtained through the measurement.
This is the physical reason why the right-hand side of Eq.~(\ref{SUwoFC}) deviates from one.
This deviation is precisely quantified by the singular probability.
Relations between nonequilibrium equalities and this kind of information loss are discussed in detail in Ref. \cite{AFMU14}.

\subsubsection{Two-particle Szilard Engine}
\begin{figure}
\begin{center}
\includegraphics[width = 1.0\columnwidth]{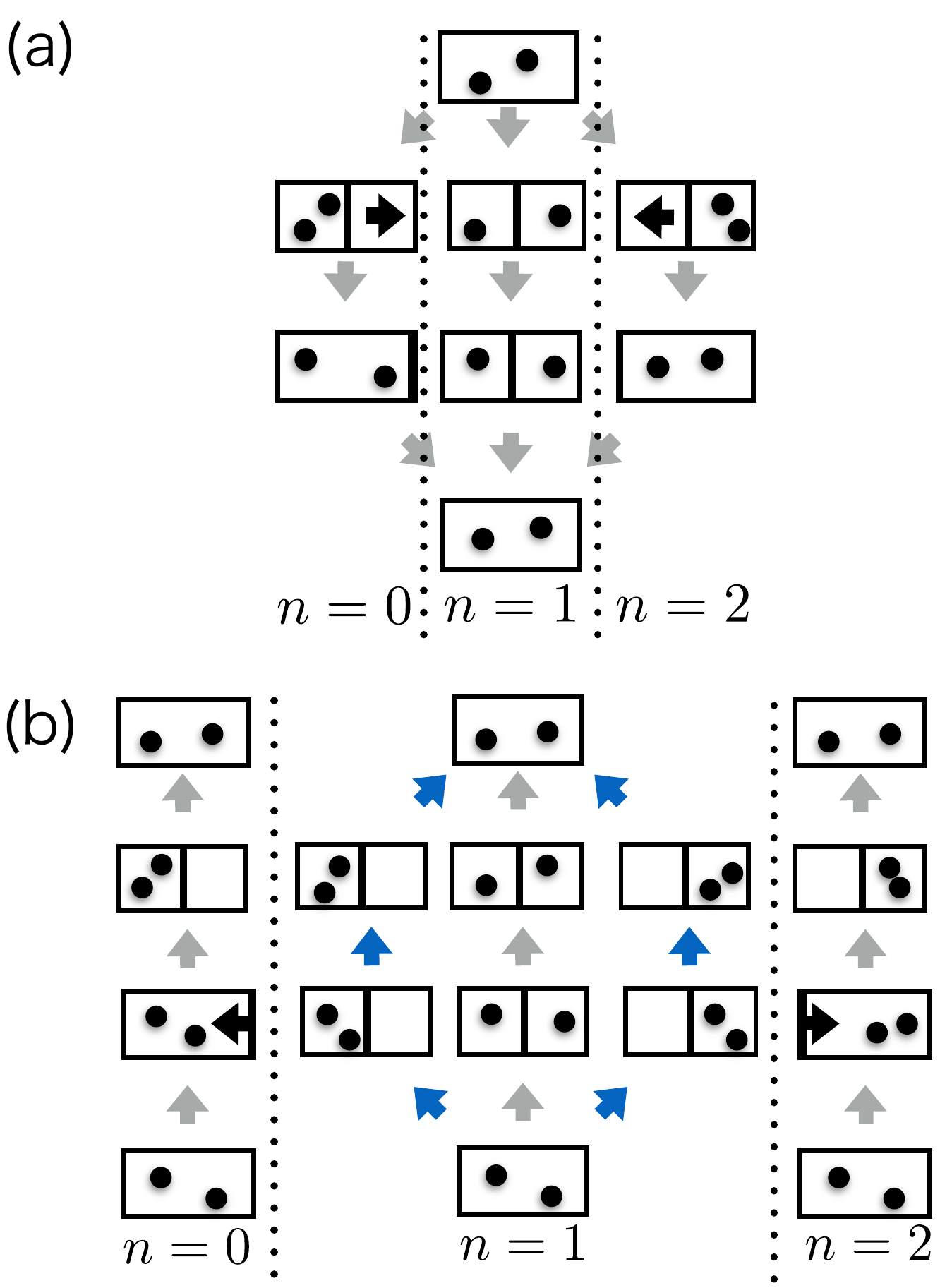}
\caption{
Two-particle Szilard engine.
(a) Forward protocol.
Initially, two particles are in thermal equilibrium.
A hard wall is inserted in the middle and a position measurement is performed to find the number $n$ of particles in the right box (second row).
Based on the measurement outcome, the wall is isothermally shifted to such a position that the extracted work is maximal (third row).
Finally, the wall is removed.
(b) Time-reversed protocol.
A wall is inserted (third row) and then moved in accordance with the value of the outcome $n$ (second row).
The case of $n=1$ contains two singular paths which are absent in the forward protocol as indicated by the blue arrows.
\label{fig:2SZE}
}
\end{center}
\end{figure}

We consider the two-particle Szilard engine to verify Eq.~(\ref{SUinAIP}).
The reason why we do not discuss the single-particle Szilard engine as in Refs. \cite{SU10, HV10} is that there arises no singular part in the single-particle Szilard engine because we can fully utilize the information obtained by the measurement and free-expansion-like dissipation does not occur in this case.
Therefore, to observe effects by absolute irreversibility, we should consider a Szilard engine with two or more particles.
Here, we consider the two-particle Szilard engine as a prototypical example.

Two indistinguishable classical-gas particles are confined in a box (Fig.~\ref{fig:2SZE}~(a)).
We insert a hard wall in the middle of the box and perform an error-free position measurement to determine the number $n$ of particles in the right box.
With probability $1/4$, we obtain the outcome $n=0$ and the mutual information $I=-\ln(1/4)$.
We then isothermally shift the wall to the right end to obtain $W=-2k_{\rm B}T\ln 2$, and finally remove the wall.
With probability $1/4$, we obtain $n=2$ and $I=-\ln(1/4)$.
In this case, we shift the wall to the opposite direction to obtain $W=-2k_{\rm B}T\ln 2$.
With probability $1/2$, we obtain $n=1$ and $I=-\ln(1/2)$.
In this case, we just remove the wall because we cannot extract any work by shifting the wall: $W=0$.
For all $n$, $\Delta F=0$ because the system returns to the initial state after the protocol.
Then, we obtain
\eqn{\label{ex}
	&&\langle e^{-\beta(W-\Delta F)-I} \rangle\nonumber\\
	&&= \frac{1}{4}e^{2\ln 2 +\ln\frac{1}{4}} + \frac{1}{2}e^{0+\ln\frac{1}{2}} +\frac{1}{4}e^{2\ln 2 +\ln\frac{1}{4}}=\frac{3}{4},
}
which deviates from $1$ because the wall removal causes free expansion when $n=1$.

Let us consider the time-reversed protocol when $n=1$
\textcolor{black}{ (Fig.~\ref{fig:2SZE}(b)).
In this protocol, we insert a wall in the middle and then remove it.
The probabilities to find the two particles in the left and right boxes are both nonzero, whereas the corresponding forward-path probabilities vanish.
Thus, the singular probability corresponding to the outcome $n=1$ is the sum of these probabilities: $\lambda_{{\rm S}|n=1}=1/2$.
On the other hand, the cases of $n=0,2$ have no singular paths:
$\lambda_{{\rm S}|n=0}=\lambda_{{\rm S}|n=2}=0$.
Therefore, we obtain
\eqn{
	\overline\lambda_{\rm S} =
	\frac{1}{4}\lambda_{{\rm S}|n=0}
	+\frac{1}{2}\lambda_{{\rm S}|n=1}
	+\frac{1}{4}\lambda_{{\rm S}|n=2} = \frac{1}{4}.
}
This value is consistent with Eqs.~(\ref{SUinAIP}) and (\ref{ex}).
}

\section{Conclusion}
We have proposed a concept of {\it absolute irreversibility} in nonequilibrium processes and show that it renders conventional integral fluctuation theorems inapplicable.
Absolute irreversibility is mathematically defined as the singularity of the reference probability measure, and physically related to divergence of entropy production.
Lebesgue's decomposition theorem enables us to generalize nonequilibrium equalities to absolutely irreversible processes since the theorem uniquely separate the absolutely irreversible part from the ordinary irreversible part.
The obtained equalities contain two physical quantities related to irreversibility:
the entropy production $\sigma$ characterizing ordinary irreversibility and the singular part probability $\lambda_{\rm S}$ describing absolute irreversibility.
In absolutely irreversible cases, the inequalities derived from our equalities are stronger than the conventional second law of thermodynamics.
We demonstrate the validity of our equalities in free expansion and the two overdamped Langevin systems.
Moreover, we generalize the nonequilibrium equalities under measurements and feedback control to processes involving absolute irreversibility caused by high-precision measurements, and illustrate the generalized equalities in two simple models.

\section*{Acknowledgement}
This work was supported by
KAKENHI Grant No. 22340114 from the Japan Society for the Promotion of Science, 
and a Grant-in-Aid for Scientific Research on Innovation Areas ``Topological Quantum Phenomena'' (KAKENHI Grant No. 22103005),
and the Photon Frontier Network Program from MEXT of Japan.
We thank Shin-ichi Sasa and Takahiro Sagawa for their critical comments.
Y. M. thanks Yui Kuramochi and Tomohiro Shitara for fruitful discussion on the mathematical aspects of our theory.
Y. M. was supported by Japan Society for the Promotion of
Science through Program for Leading Graduate Schools (MERIT).
K. F. was supported by Japan Society for the Promotion of
Science through Advanced Leading Graduate Course for Photon Science (ALPS).
K. F. acknowledges support from JSPS (Grant No. 254105).

\begin{appendix}
\section{Choice of the Reference Probability Measure and Entropy Production}
In this appendix, we will show that a proper choice of the reference probability measure results in the corresponding nonequilibrium equality with physical and experimentally observable entropy production.
We explicitly deal with a Langevin system and a Hamiltonian system.
Here, we consider for simplicity cases in the absence of feedback control.
The extension to cases under feedback control is straightforward.

\subsection{Langevin System}
To elucidate the physical meaning of Eq.~(\ref{Jar}), 
let us first consider an overdamped Langevin system with white Gaussian noise:
\eqn{
	\dot x(t) = \mu F(x(t), \lambda(t)) + \zeta(t),
}
where $\mu$, $F$ and $\lambda$ are the mobility, systematic force, and control parameter, respectively, and $\zeta(t)$ is a random noise satisfying $\langle \zeta(t)\zeta(s) \rangle = 2D\delta (t-s)$, where $D$ is diffusion constant.
The path probability generated by this dynamics is \cite{Sei12, Sekimoto}
\eqn{
	\mM[\mD x] = \mathcal{N}e^{-\mathcal{A}[x, \lambda]}\mu_0(dx_0)\mD\tilde x \mu_{\rm L}(dx_\tau),
}
where 
\eqn{
	\mathcal{A}[x, \lambda] =
	\int_0^\tau dt \left[
		\frac{(\dot x -\mu F)^2}{4D}+\frac{\mu}{2}\frac{\partial F}{\partial x}
	\right],
}
and $\mathcal{N}$, $\mu_0$ and $\mu_{\rm L}$ represent the normalization constant, the initial probability distribution measure and the Lebesgue measure on the configuration space $\Omega$, respectively.
Moreover, $x_0$ and $x_\tau$ represent the initial and final positions of path $x$, respectively, and $\tilde x$ is the path that does not have these endpoints.
To quantify the asymmetry under time reversal, we also consider the realization probability of the time-reversed path:  $x^\dag(t)=x(\tau-t)$.
The time-reversed probability is
\eqn{
	\mM^\dag[\mD x^\dag]&=&\mathcal{N} e^{-\mathcal{A}[x^\dag,\lambda^\dag]}\mu_0^\dag(dx_0^\dag)\mD \tilde x^\dag \mu_{\rm L}(dx_\tau^\dag)\\
	&=& \mathcal{N} e^{-\mathcal{A}[x^\dag,\lambda^\dag]}
	\mu_{\rm L}(dx_0)\mD\tilde x \mu_0^\dag(dx_\tau),
}
where $\mu_0^\dag$ is the initial probability measure of the time-reversed process.
Let $\mu_{L,{\rm AC}}$ denote the absolutely continuous part of $\mu_{\rm L}$ with respect to $\mu_0$, and let $\mu^\dag_{0,{\rm AC}}$ denote the absolutely continuous part of $\mu_0^\dag$ with respect to $\mu_{\rm L}$.
Then, the absolutely continuous part of $\mM^\dag$ with respect to $\mM$ is
\eqn{\label{LanAC}
	\mM_{\rm AC}^\dag[\mD x^\dag]=
	\mathcal{N} e^{-\mathcal{A}[x^\dag,\lambda^\dag]}
	\mu_{L,{\rm AC}}(dx_0)\mD\tilde x \mu_{0,{\rm AC}}^\dag(dx_\tau).\ \ \ \ \ \ \
}
Therefore, the entropy production is defined as
\eqn{
	e^{-\sigma} 
	&=& \left.\frac{\mD\mM^\dag_{\rm AC}[\mD x^\dag]}{\mD\mM[\mD x]}\right|_x\nonumber\\
	&=& e^{-(\mA[x^\dag,\lambda^\dag]-\mA[x,\lambda])}
	\left.\frac{d\mu_{L, {\rm AC}}}{d\mu_0}\right|_{x_0}\hspace{-5pt}
	\left.\frac{d\mu^\dag_{0,{\rm AC}}}{d\mu_L}\right|_{x_\tau}\hspace{-5pt}.
}
By a simple calculation, we can verify
\eqn{
	\mathcal{A}[x^\dag, \lambda^\dag]
	= \int_0^\tau dt \left[
		\frac{(\dot x +\mu F)^2}{4D}+\frac{\mu}{2}\frac{\partial F}{\partial x}
	\right],
}
and therefore
\eqn{
	\mA[x^\dag,\lambda^\dag]-\mA[x,\lambda]
	&=& \frac{\mu}{D}\int_0^\tau dt\dot x F\nonumber.
}
Considering the Einstein relation $D=\mu k_BT$, we obtain
\eqn{
	\mA[x^\dag,\lambda^\dag]-\mA[x,\lambda] = \beta\int_0^\tau dt \dot x F,
}
where $\beta$ is the inverse temperature.
The integral in the right-hand side is the work done by the systematic force.
Since this work is instantly dissipated to the heat bath by the assumption of overdamping, the work is equal to dissipated heat $Q$.
Therefore, we obtain
\eqn{
	\mA[x^\dag,\lambda^\dag]-\mA[x,\lambda] = \beta Q,
}
and
\eqn{\label{defsig}
	e^{-\sigma} =
	e^{-\beta Q}
	\left.\frac{d\mu_{L, {\rm AC}}}{d\mu_0}\right|_{x_0}
	\hspace{-5pt}
	\left.\frac{d\mu^\dag_{0,{\rm AC}}}{d\mu_L}\right|_{x_\tau}
	\hspace{-5pt}.
}
It should be stressed that the dissipated heat defined here is the standard definition in mesoscopic systems \cite{Sek97, Sekimoto, Sei12}.
Moreover, this heat is an observable quantity in experiments as is measured in Refs.~\cite{SBBS07,JPC08} for example.

\subsubsection{Dissipated Work}
Here, let us assume that the initial state is a local equilibrium state in a region $D_0\subset\Omega$:
\eqn{
	\mu_0(dx_0) = e^{-\beta(U(x_0, \lambda_0)-F_0)}\chi_{D_0}(x_0)\mu_L(dx_0),
}
where $U$ is the potential and $\chi_{D_0}(x_0)$ is the characteristic function defined by
\eqn{
	\chi_{D_0}(x_0)=
	\left\{
	\begin{array}{ll}
		1 & x_0\in D_0\\
		0 & x_0\notin D_0,
	\end{array}
	\right.
}
and $F_0$ is defined by
\eqn{
	e^{-\beta F_0} = \int_{D_0} e^{-\beta U(x_0,\lambda_0)} d\mu_L(x_0).
}
In this case, the absolutely continuous part of $\mu_L$ with respect to $\mu_0$ is
\eqn{
	\mu_{L,{\rm AC}}(dx_0) = \chi_{D_0}(x_0)\mu_L(dx_0)
}
and therefore
\eqn{\label{rd1}
	\left.\frac{d\mu_{L,{\rm AC}}}{d\mu_0}\right|_{x_0} = e^{\beta(U(x_0, \lambda_0)-F_0)}.
}

To proceed further, let us set the initial probability measure of the time-reversed dynamics to a local equilibrium distribution in a region $D_\tau$:
\eqn{
	\mu_0^\dag(dx_\tau) =
	e^{-\beta(U(x_\tau,\lambda_\tau)-F_\tau)}
	\chi_{D_\tau}(x_\tau)\mu_L(dx_\tau),
}
which is already absolutely continuous with respect to $\mu_L$, {\it i.e.}, $\mu_{0,{\rm AC}}^\dag = \mu_0^\dag$.
Thus, we obtain
\eqn{\label{rd2}
	\left.\frac{d\mu^\dag_{0,{\rm AC}}}{d\mu_L}\right|_{x_\tau}
	=e^{-\beta(U(x_\tau,\lambda_\tau)-F_\tau)}\chi_{D_\tau}(x_\tau).
}
Substituting Eqs.~(\ref{rd1}) and (\ref{rd2}) into Eq.~(\ref{defsig}), we obtain
\eqn{
	e^{-\sigma} = e^{-\beta(Q+\Delta U - \Delta F)}\chi_{D_\tau}(x_\tau),
}
where $\Delta U=U(x_\tau,\lambda_\tau)-U(x_0,\lambda_0)$ and $\Delta F=F_\tau-F_0$.
Therefore, when $x_\tau\in D_\tau$,
\eqn{
	\sigma &=& \beta(Q+\Delta U-\Delta F)\\
	&=& \beta(W-\Delta F),
}
where we use the first law of thermodynamics: $W=Q+\Delta U$.
Since the right-hand side is a part of work which is not converted to free energy, it is called dissipated work.
When $x_\tau\notin D_\tau$, $\sigma$ is positively divergent.
(Positively divergent entropy production does not cause any problem in the context of integral fluctuation theorems since it contributes none to the exponentiated average.)
Thus, Eq.~(\ref{Jar}) reduces to
\eqn{\label{DWRinAIP}
	\langle \chi_{D_\tau}(x_\tau)e^{-\beta(W-\Delta F)}\rangle = 1-\lambda_{\rm S}.
}
In particular, when we set $D_\tau$ to $\Omega$, we have
\eqn{\label{JinAIP}
	\langle e^{-\beta(W-\Delta F)}\rangle = 1-\lambda_{\rm S}.
}
This is an extension of the Jarzynski equality \cite{Jar97Apr, Jar97Nov}.

\subsubsection*{Application to a Specific Example}
Let us apply Eq.~(\ref{JinAIP}) to the example in Sec. II C 2.
In this case, $x(t)$ is the coordinate along the ring.
The half width of a single well is defined as $a$.
Therefore, $x(t)$ has the periodic boundary condition: $x(t)=x(t)+2na$.
Here, $D_0$ is the region of the initial well of the original process and $D_\tau$ is the entire configuration space $\Omega$.
Therefore, from Eq.~(\ref{LanAC}), we have
\eqn{
	&&\mM^\dag_{\rm AC}[\mD x^\dag]\nonumber\\
	&&=\mN e^{-\mA[x^\dag,\lambda^\dag]}\chi_{D_0}(x_0)\mu_L(dx_0)\mD\tilde x \mu^\dag_0(dx_\tau).\ \ \ 
}
Let us calculate the singular probability $\lambda_{\rm S}$ by utilizing the $n$-fold symmetry of the system.
As the potential has the fundamental period $2a$, we have
\eqn{
	F(x(t)+2ma,\lambda(t)) &=& F(x(t),\lambda(t)),\label{tt1}\\
	\mu^\dag_0(d(x_\tau+2ma)) &=& \mu^\dag_0(dx_\tau),\label{tt2}\\
	\mA[x^\dag+2ma,\lambda^\dag] &=& \mA[x^\dag,\lambda^\dag]\label{tt3},
}
for an arbitrary integer $m$.
Let $D_0^m$ denote the region of the $m$-th well from the initial well, and we have
\eqn{\label{tt4}
	\chi_{D_0^m}(x_0+2ma)=\chi_{D_0}(x_0) .
}
As the integration is invariant under spatial translation, from Eqs.~(\ref{tt1}), (\ref{tt2}), (\ref{tt3}) and (\ref{tt4}), we obtain
\eqn{
	&&\int \mM^\dag_{\rm AC}[\mD x^\dag]\nonumber\\
	&&=\int \mM^\dag_{\rm AC}[\mD (x^\dag+2ma)]\nonumber\\
	&&=\int\mN e^{-\mA[x^\dag,\lambda^\dag]}\chi_{D_0^{-m}}(x_0)\mu_L(dx_0)\mD\tilde x\mu_0^\dag(dx_\tau).\ \ \ 
}
Therefore, we have
\eqn{
	&&n\int \mM^\dag_{\rm AC}[\mD x^\dag]\nonumber\\
	&&=\sum_{m=0}^{n-1}\int\mN e^{-\mA[x^\dag,\lambda^\dag]}\chi_{D_0^m}(x_0)\mu_L(dx_0)\mD\tilde x\mu_0^\dag(dx_\tau)\nonumber\\
	&&=\int\mN e^{-\mA[x^\dag,\lambda^\dag]}\sum_{m=0}^{n-1}\chi_{D_0^m}(x_0)\mu_L(dx_0)\mD\tilde x\mu_0^\dag(dx_\tau)\nonumber\\
	&&=\int\mN e^{-\mA[x^\dag,\lambda^\dag]}\mu_L(dx_0)\mD\tilde x\mu_0^\dag(dx_\tau)\nonumber\\
	&&=\int \mM^\dag[\mD x^\dag]\nonumber\\
	&&=1.
}
Thus, we obtain
\eqn{
	\lambda_{\rm AC} = \frac{1}{n},\ \lambda_{\rm S} = \frac{n-1}{n}.
}
Equation~(\ref{JinAIP}) reduces to
\eqn{
	\langle e^{-\beta(W-\Delta F)} \rangle = \frac{1}{n}.
}

\subsubsection{Total Entropy Production}
Here, we assume that the initial probability distribution of the original process can be written in terms of a probability density, that is, $\mu_0$ is absolutely continuous with respect to $\mu_L$:
\eqn{
	\mu_0(dx_0) = p_0(x_0)\mu_L(dx_0).
}
The absolutely continuous part of $\mu_L$ with respect to $\mu_0$ is
\eqn{
	\mu_{L, {\rm AC}}(dx_0)=\chi_{D_0}(x_0)\mu_L(dx_0),
}
where $D_0$ is the support of $p_0$.
Therefore,
\eqn{\label{rd3}
	\left.\frac{d\mu_{L,{\rm AC}}}{d\mu_0}\right|_{x_0}
	 = \frac{\chi_{D_0}(x_0)}{p_0(x_0)}.
}
Let us set the initial probability distribution of the time-reversed process to be equal to the final probability distribution of the original process: $\mu^\dag_0 = \mu_\tau$.
Since $\mu^\dag_{0,{\rm AC}}$ is absolutely continuous with respect to the Lebesgue measure, we obtain
\eqn{\label{rd4}
	\left.\frac{d\mu^\dag_{0,{\rm AC}}}{d\mu_L}\right|_{x_\tau}
	= p_\tau(x_\tau),
}
where $p_\tau$ is an (unnormalized) integrable function.
Substituting Eqs.~(\ref{rd3}) and (\ref{rd4}) into Eq.~(\ref{defsig}), we have
\eqn{
	e^{-\sigma} = e^{-\Delta s^{\rm bath}}\frac{p_\tau(x_\tau)}{p_0(x_0)}\chi_{D_0}(x_0),
}
where $\Delta s^{\rm bath}=\beta Q$ is the entropy production of the heat bath.
When $x_0\in D_0$, we obtain
\eqn{
	\sigma = \Delta s^{\rm bath}-\ln p_\tau(x_\tau)+\ln p_0(x_0).
}
Defining the (unavaraged) Shannon entropy $s(t)=-\ln p_t(x_t)$, we obtain
\eqn{
	\sigma = \Delta s^{\rm bath} + \Delta s,
}
where $\Delta s=s(\tau)-s(0)$ is the difference in the Shannon entropy.
In this case, $\sigma$ is the sum of the Shannon entropy production of the system and the thermodynamic entropy production of the bath, namely total entropy production $\Delta s^{\rm tot}$.
Thus, Eq.~(\ref{Jar}) reduces to
\eqn{\label{TERinAIP}
	\langle e^{-\Delta s^{\rm tot}} \rangle = 1-\lambda_{\rm S}.
}

\subsection{Hamiltonian System}
In a Hamiltonian system, the dynamics is deterministic.
Therefore, the probability of a given path is the same as the probability of the system being found on the same path at an arbitrary time.
In other words, the probability measures can be replaced as follows:
\eqn{
	\mM[\mD\Gamma] &=& \mu_{0}(d\Gamma_{0})\mD\tilde \Gamma \mu_L(d\Gamma_\tau),\\
	\mM^\dag[\mD\Gamma^\dag] &=& \mu_0^\dag(d\Gamma_0^\dag)\mD\tilde\Gamma^\dag \mu_L(d\Gamma_\tau^\dag)\nonumber\\
	&=& \mu_L(d\Gamma_0)\mD\tilde\Gamma\mu_0^\dag(d\Gamma_\tau).
}
Let us separate the degrees of freedom into two parts;
one involves the degrees of freedom $x$ concerning the system of interest and the other involves the degrees of freedom  $y$ concerning the bath.
We also assume that the Hamiltonian of the bath $H^{\rm bath}(y)$ is independent of time.
Initially, the bath is in equilibrium:
\eqn{
	\mu_0(d\Gamma_0) = \mu^{\rm sys}_0(dx_0)e^{-\beta(H^{\rm bath}(y_0)-F^{\rm bath})}\mu_L^{\rm bath}(dy_0),\ \ \ \ \ \ \ 
}
where $\mu_0^{\rm sys}$ and $\mu_L^{\rm bath}$ represent the  initial probability distribution measure of the system and the Lebesgue measure of the bath, respectively, and
\eqn{
	e^{-\beta F^{\rm bath}} = \int dy e^{-\beta H^{\rm bath}(y)}.
}
Moreover, let us set the initial probability measure of the time-reversed process to the product of the initial probability measure of the system $\mu_0^{\rm sys\dag}$ and the canonical probability measure of the bath:
\eqn{
	&&\mu_0^\dag(d\Gamma_\tau)\nonumber\\
	&&=\mu_0^{\rm sys\dag}(dx_\tau) e^{-\beta(H^{\rm bath}(y_\tau)-F^{\rm bath})}\mu_L^{\rm bath}(dy_\tau).
}
Therefore, we obtain
\eqn{
	e^{-\sigma}
	&=& \frac{\mD\mM^\dag_{\rm AC}[\mD\Gamma^\dag]}{\mD\mM[\mD\Gamma]}\nonumber\\
	&=& e^{-\beta(H^{\rm bath}(y_\tau)-H^{\rm bath}(y_0))}\hspace{-5pt}
	\left.\frac{d\mu_{L,{\rm AC}}^{\rm sys}}{d\mu^{\rm sys}_0}\right|_{x_0}\hspace{-5pt}
	\left.\frac{d\mu^{\rm sys\dag}_{0,{\rm AC}}}{d\mu_L^{\rm sys}}\right|_{x_\tau}\hspace{-8pt}.\ \ \ \ \ \ \ 
}
Since the increase in bath's energy is heat dissipated from the system, we have
\eqn{
	H^{\rm bath}(y_\tau)-H^{\rm bath}(y_0) = Q.
}
Therefore, we obtain
\eqn{\label{Hamexpsig}
	e^{-\sigma}
	= e^{-\beta Q}
	\left.\frac{d\mu_{L,{\rm AC}}^{\rm sys}}{d\mu^{\rm sys}_0}\right|_{x_0}
	\left.\frac{d\mu^{\rm sys\dag}_{0,{\rm AC}}}{d\mu_L^{\rm sys}}\right|_{x_\tau},
}
where $\mu^{\rm sys}_{L,{\rm AC}}$ is the absolutely continuous part of $\mu^{\rm sys}_L$ with respect to $\mu^{\rm sys}_0$, and $\mu^{\rm sys\dag}_{0,{\rm AC}}$ is the absolutely continuous part of $\mu^{\rm sys\dag}_0$ with respect to $\mu^{\rm sys}_L$.
Since Eq.~(\ref{Hamexpsig}) has exactly the same structure as that of Eq.~(\ref{defsig}), it is obvious that Eqs.~(\ref{DWRinAIP}) and (\ref{TERinAIP}) can be derived under the same assumptions and choices of the reference probability in this Hamiltonian system.

\section{Comparison with Conventional Methods}
In this section, we review conventional methods \cite{Sasa} to compare with our method described in the main text.

\subsection{Rederivation of Eq.~(\ref{C1ave})}
\begin{figure}
\begin{center}
\includegraphics[width = 0.25\columnwidth]{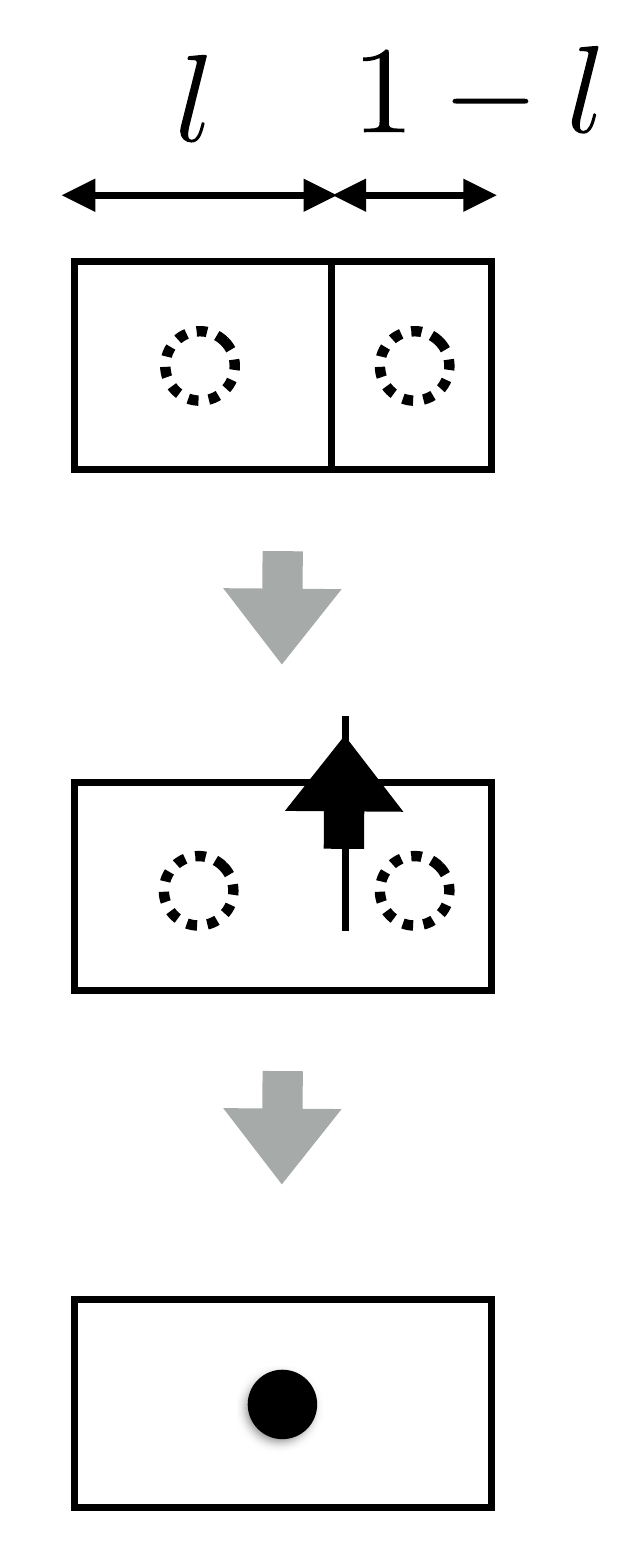}
\caption{
Virtual process corresponding to free expansion.
Initially, the particle is in the global equilibrium state;
the particle is in the left box with probability $l$ and in the right one with probability $1-l$.
Then, the wall is removed.
\label{fig:VFE}
}
\end{center}
\end{figure}
Here, we rederive Eq.~(\ref{C1ave}) according to the conventional method.
We consider a virtual process starting from the global equilibrium state.
The particle is in the left box with probability $l$ and in the right one with probability $1-l$ so that the probability distribution is globally uniform.
Since the initial probability distribution vanishes nowhere in phase space, there arises no absolute irreversibility.
Therefore, the conventional nonequilibrium equality applies \cite{Cro00}:
\eqn{\label{VJE}
	[\mF[\Gamma]e^{-\beta(\hat W[\Gamma] - \Delta\hat F)}]
	=[\mF^\dag[\Gamma^\dag]]^\dag,
}
where $\mF[\Gamma]$ is an arbitrary path-dependent functional, and $\mF^\dag$ is defined by $\mF^\dag[\Gamma^\dag]=\mF[\Gamma]$;
$[\cdots]$ denotes the statistical average in the virtual process, and the symbol $\hat{}$ means that the accompanying quantity is the one in the virtual process.
In this simple case, we can straightforwardly connect the physical quantities in the virtual process and those in the original one thorough the following relations:
\eqn{
	\hat W[\Gamma] &=& W[\Gamma],\label{VW}\\
	\hat F_0 &=& F_0 + k_BT \ln l,\\
	\hat F_\tau &=& F_\tau,\\
	\Delta \hat F &=& \Delta F - k_BT \ln l\label{VF}.
}
Moreover, let us define $\mF[\Gamma]$ as the characteristic functional whose value is equal to one only if the path starts from the left box and zero otherwise.
Then, considering the normalization of probabilities properly, we can connect the average in the virtual process to that of the original one as
\eqn{
	[\mF[\Gamma]\cdots] = l\langle \cdots \rangle \label{Vave}
}
Finally, in the time-reversed process, probability of paths ending in the left box is $l$, and therefore we have
\eqn{
	[\mF^\dag[\Gamma^\dag]]^\dag = l.\label{VRHS}
}
By substituting Eqs.~(\ref{VW}), (\ref{VF}), (\ref{Vave}) and (\ref{VRHS}) into Eq.~(\ref{VJE}), we can reproduce Eq.~(\ref{C1ave}).

Following the same procedure, we can also rederive Eqs.~(\ref{sc}) and (\ref{d}).
In the next section, we discuss the meaning of this conventional derivation and compare it with our method based on the Lebesgue decomposition.

\subsection{Interpretation and Discussion}
Let us examine the meaning of the conventional procedure.
First of all, we extend the initial probability distribution to the canonical distribution of the entire system to avoid the problem of the vanishing probability.
Then, we derive the nonequilibrium equality in this artificial process (Eq. (\ref{VJE})).
Next, we express a physical quantity of the artificial process by that of the original process (Eqs.~(\ref{VW}) and (\ref{VF})).
Finally, we erase some of paths to relate the average of the original system to that of the artificial system (Eq.~(\ref{VRHS})), and obtain the nonequilibrium equality in the original system (Eq.~(\ref{C1ave})).

In this way, the conventional method has to introduce the artificial system to derive the nonequilibrium equality in the absolutely irreversible process; in contrast our method, based on Lebesugue's decomposition, directly deals with the original process.
Moreover, the reason why the right-hand sides of Eqs.~(\ref{C1ave}), (\ref{sc}) and (\ref{d}) deviate from one is crystal-clear in our method, that is, the probabilities of absolutely irreversible paths are subtracted from one.
Incidentally, we note that although Eq.~(\ref{d}) can in hindsight be derived by the conventional method, we naturally find this absolutely irreversible example by Lebesgue's decomposition theorem (17).

We also note that we can derive Eqs.~(\ref{C1ave}) and (\ref{sc}) by partitioning the phase space into the part in which the initial probability vanishes and the part in which the initial probability dose not vanish before we consider the time evolution of the system \cite{Sagawa},
by utilizing the method used to survey the coarse-grained property of the Kullback-Leibler divergence \cite{KPvB07},
that is, we can obtain Eqs.~(\ref{C1ave}) and (\ref{sc}) from Eq.~(7) in Ref. \cite{KPvB07}.

Let us turn to consider cases with feedback control.
In these cases, to extend the initial probability to the entire system is not enough to eliminate absolutely irreversiblity,
since high-precision measurements can forbid some forward paths for a given outcome.
Therefore, the conventional methods do not work straightforwardly.
However, we can consider virtual measurements with enough errors that all forward paths are allowed for a given outcome and remove absolute irreversibility.
In this virtual process with erroneous measurements, we can safely derive the conventional integral fluctuation theorem.
Then, we connect physical quantities in the virtual process to those in the original process and erase the irrelevant paths by a properly-chosen filter function, in a manner similar to cases without feedback control, and obtain nonequilibrium equality in the original process.

In summary, although Eqs.~(\ref{Jar}) and (\ref{SU}) can be derived by the conventional methods in principle, our method is simpler in that processes of interest can be directly treated, and moreover provides a clearer meaning of the obtained nonequilibrium equality.
\end{appendix}

\bibliography{reference}
\end{document}